\documentclass[12pt,a4paper]{article}
\usepackage{mathrsfs}
\usepackage{epsfig}
\begin{document}
\title{Single production of the doubly charged $Higgs$ boson via
$e\gamma$ collision in the $Higgs$ triplet model}
\author{Chong-Xing Yue, Xue-Song Su, Jiao Zhang, Jue Wang \\
{\small Department of Physics, Liaoning  Normal University, Dalian
116029, P. R. China}
\thanks{E-mail:cxyue@lnnu.edu.cn}}
\date{\today}

\maketitle
\begin{abstract}
The $Higgs$ triplet model ($HTM$) predicts the existence of a pair
of doubly charged $Higgs$ bosons $H^{\pm\pm}$. Single production of
$H^{\pm\pm}$ via $e\gamma$
 collision at the next generation
$e^{+}e^{-}$ International Linear Collider ($ILC$) and the  Large
Hadron electron Collider ($LHeC$) is considered. The numerical
results show that the production cross sections are very sensitive
to the neutrino oscillation parameters. Their values for the
inverted hierarchy mass spectrum are larger than those for the
normal hierarchy mass spectrum at these two kinds of collider
experiments. With reasonable values of the relevant free parameters,
the possible signals of the  doubly charged $Higgs$ bosons predicted
by  the $HTM$ might be detected in future $ILC$ experiments.

\vspace{1cm}

\vspace{2.0cm} \noindent {\bf PACS number}:  12.60.Cn, 14.80.Cp,
13.10.+q

\end{abstract}
\newpage
\noindent{\bf 1. Introduction}

During the past decade, neutrino oscillation experiments have
provided us with very convincing evidence that neutrinos are massive
particles mixing with each other [1]. This exciting breakthrough
gives a strong motivation for physics beyond the standard model
($SM$), since the $SM$ itself only contains three massless neutrinos
[2]. Various ways to go beyond the $SM$ have been proposed in order
to accomodate this observation. The $Higgs$ triplet model ($HTM$)
[3, 4] is a simple extension of the $SM$ with an $SU(2)_{L}$-triplet
$Higgs$ boson of hypercharge $Y=2$ whose vacuum expectation value
($VEV$) $\nu_{\triangle}$ provides Majorana neutrino masses without
introducing right-handed neutrinos.

The $HTM$ [3, 4] generates a Majorana neutrino mass via the product
of a triplet $Yukawa$ coupling $h_{ij}$ and a triplet $VEV$
$\nu_{\triangle}$. This model predicts the existence of seven
physical $Higgs$ bosons which are two $CP$-even neutral bosons
$h^{0}$ and $H^{0}$, a $CP$-odd neutral one $A^{0}$, a pair of
single charged bosons $H^{\pm}$, and a pair of doubly charged
$Higgs$ bosons $H^{\pm\pm}$. These new particles can produce rich
phenomenology at present and in future high energy collider
experiments. Since charge conservation prevents doubly charged
scalars from decaying to a pair of quarks, $H^{\pm\pm}$ carry the
lepton number violation and would give rise to a distinctive
experimental signal, which has lower background. This fact has lead
to many studies involving the doubly charged $Higgs$ bosons
$H^{\pm\pm}$ in the literature. For example, it has been shown that
such particles can be produced with sizable rates at hadron
colliders via the processes $q\bar{q}\rightarrow H^{++}H^{--}$ [5,
6] and $q\bar{q'}\rightarrow H^{\pm\pm}H^{\mp}$ [5, 7, 8]. Using
these production mechanisms, discovery of $H^{\pm\pm}$ at the Large
Hadron Collider ($LHC$) has been studied in Refs. [9, 10].

There is a direct connection between the $Yukawa$ coupling $h_{ij}$
and the neutrino mass matrix which gives rise to phenomenological
prediction for processes which depend on $h_{ij}$ [11]. The
branching ratios of the lepton flavor violating ($LFV$) decay
processes $l_{i}\rightarrow l_{j}l_{l}l_{k}$ and $l_{i}\rightarrow
l_{j}\gamma$ contributed by the doubly charged $Higgs$ bosons
$H^{\pm\pm}$ are strong depend on $h_{ij}$ in the context of the
$HTM$. Recently, it has been shown [12] that the current upper
experimental limits for some of these $LFV$ decay processes can give
serve constraints on the parameter space of $h_{ij}$ in the $HTM$.
Taking into account these constraints, in this paper, we consider
single production of $H^{\pm\pm}$ via $e\gamma$ collision at the
next generation $e^{+}e^{-}$ International Linear Collider ($ILC$)
[13, 14] and the Large Hadron electron Collider ($LHeC$) [15].
Although single production of the doubly charged $Higgs$ bosons via
$e\gamma$ collision has been studied in Refs. [16], in this paper,
we will focus our attention on the effects of the neutrino
oscillation parameters on its production cross section and see
whether the neutrino mass hierarchy can be disentangled via this
kind of production processes at the $ILC$ and $LHeC$. Our numerical
results show that the production cross sections at these two kinds
of high energy colliders are very sensitive to the neutrino
oscillation parameters. The production cross section of the process
$e\gamma\rightarrow e^{+}H^{--}$ for the inverted hierarchy mass
spectrum is significantly larger than that for the normal hierarchy
mass spectrum. In the sizable range of the parameter space of the
$HTM$, their values can be large enough to be detected in the future
$ILC$ experiments.

Our paper is organized as follows. The essential features of the
$HTM$ are briefly reviewed in section 2. In this section, we also
summarized the constraints on the $HTM$ from some $LFV$ decay
processes. The effective production cross sections for the
subprocess $e\gamma\rightarrow l^{+}H^{--}$ at the $ILC$ and $LHeC$
are calculated in sections 3 and 4. Our conclusions and discussions
are given in section 5.

\noindent{\bf 2. The essential features of the $\bf HTM$}

The $Higgs$ triplet model ($HTM$) [3, 4] is one of the appealing
scenarios for generating neutrino mosses without the introduction of
a right-handed neutrino. In this model, a complex $SU(2)_{L}$
triplet scalar with the hypercharge $Y=2$ is introduced to the $SM$,
which can be parameterized by

\begin{equation}
\triangle=\left(\begin{array}{cc}\triangle^{+}/\sqrt{2}&\triangle^{++}\\\triangle^{0}&\
-\triangle^{+}/\sqrt{2}
\end{array}\right).
\end{equation}
A nonzero $VEV$ value $<\triangle^{0}>$ gives rise to the following
mass matrix for neutrinos

\begin{equation}
m_{ij}=2h_{ij}<\triangle^{0}>=\sqrt{2}h_{ij}\nu_{\triangle}.
\end{equation}
The symmetric complex matrix $h_{ij}$ $(i,j=e,\mu,\tau)$ is the
$Yukawa$ coupling strength. The interaction of the triplet scalar
with lepton doublet $L_{i}\equiv(\nu_{iL},l_{iL})^{T}$ is given by

\begin{equation}
\pounds=-h_{ij}[\nu_{iL}^{T}C\nu_{jL}\triangle^{0}-\frac{1}{\sqrt{2}}(\nu_{iL}^{T}C\l_{jL}+\l_{iL}^{T}C\nu_{jL})
\triangle^{+}-\l_{iL}^{T}C\l_{jL}\triangle^{++}]
             +h.c.,
\end{equation}
here $C$ is the charge conjugation operator.

The neutrino mass matrix $m_{ij}$ can be diagonalized by the
Maki-Nakagawa-Sakata ($MNS$) matrix $V_{MNS}$ [17]. Then the
couplings $h_{ij}$ can be written as following

\begin{equation}
h_{ij}=\frac{m_{ij}}{\sqrt{2}\nu_{\triangle}}\equiv\frac{1}{\sqrt{2}\nu_{\triangle}}
[V_{MNS}diag(m_{1},m_{2}e^{i\varphi_{1}},m_{3}e^{i\varphi_{2}})V_{MNS}^{T}]_{ij}
\end{equation}
with\begin{equation}
V_{MNS}=\left(\begin{array}{ccc}c_{12}c_{13}&c_{13}s_{12}&e^{-i\delta}s_{13}
\\-c_{12}s_{13}s_{23}e^{i\delta}-c_{23}s_{12}&c_{12}c_{23}-e^{i\delta}s_{12} s_{13}s_{23}&c_{13}s_{23}
\\s_{12}s_{23}-e^{i\delta}c_{12}c_{23}s_{13}&-c_{23}s_{12}s_{13}e^{i\delta}-c_{12}s_{23}&c_{13}c_{23}
\end{array}\right)\times
diag(e^{i\varphi_{1}/2},1,e^{i\varphi_{2}/2}).
\end{equation}
Where $m_{1}$, $m_{2}$, $m_{3}$ are the absolute masses of the three
neutrinos. The phase $\delta $ is the Dirac CP-violating phase,
$\varphi_{1}$ and $\varphi_{2}$ are referred to as the Majorana
phases [4, 18] and there are $0\leq \varphi_{1}$,
$\varphi_{2}<2\pi$. $s_{ij}=\sin\theta_{ij}$,
$c_{ij}=\cos\theta_{ij}$ with $0\leq\theta_{ij} \leq \pi/2$, in
which $ \theta_{ij}$ are the mixing angles. The explicit expression
forms for $h_{ij}$ have been given in Refs. [8, 9].

After the eletroweak
symmetry breaking, there are seven physical massive $Higgs$ bosons
($H^{\pm\pm}, H^{\pm}, h^{0}, H^{0}, A^{0}$) left in the spectrum
for the $HTM$. The doubly charged $Higgs$ bosons $H^{\pm\pm}$ are
entirely composed of the triplet scalars $\bigtriangleup^{\pm\pm}$,
while the remaining eigen states are mixtures of the doublet and
triplet scalars. Such mixing is proportional to the triplet $VEV$
$\nu_{\triangle}$, which is generally small even if
$\nu_{\triangle}$ assumes its largest value of a few GeV. Thus,
$H^{\pm}$, $H^{0}$, $A^{0}$ are predominantly composed of the
triplet scalars, while the $SM$-like $Higgs$ boson $h^{0}$ is mainly
composed of the doublet scalars.

It is well known that the doubly charged $Higgs$ bosons $H^{\pm\pm}$
have contributions to the $LFV$ decays $l_{i}\rightarrow
l_{j}\gamma$, $l_{j}\rightarrow l_{j}l_{k}l_{l}$ [19]. In the $HTM$,
their branching ratios depend on the neutrino mass matrix
parameters, the $VEV$ value $\nu _{\triangle}$, and the mass
$m_{H^{\pm\pm}}$ of the doubly charged Higgs bosons [12]. The
current experimental upper limits for some of these $LFV$ decay
processes can give severe constraints on the relevant free
parameters. In the $HTM$, the $LFV$ decay process $\mu\rightarrow
e\gamma$ proceed via exchange of $H^{\pm\pm}$ and $H^{\pm}$ at
one-loop. In the case of ignoring the electron mass in the final
state and all lepton masses in the loop, the branching ratio
$Br(\mu\rightarrow e\gamma)$ can be written as [12]
\begin{equation}
Br(\mu\rightarrow e\gamma)\approx
\frac{27\alpha\mid(h^{+}h)_{e\mu}\mid^{2}}{64\pi
G_{F}^{2}m^{4}_{H}},
\end{equation}
where $\alpha=1/137$ is the fine structure constant,
$G_{F}=1.17\times10^{-5}GeV^{-2}$ is the Fermi constant. In above
equation, it has been assumed $m_{H^{\pm\pm}}=m_{H^{\pm}}=m_{H}$.
The current experimental upper limit for the $LFV$ process
$\mu\rightarrow e\gamma$ is $Br^{exp}(\mu\rightarrow
e\gamma)\leq1.2\times10^{-11}$ [20], then we have

\begin{equation}
 \mid(h^{+}h)_{e\mu}\mid^{2}\leq2.6\times10^{-9}(\frac{m_{H}}{200GeV})^{4}.
\end{equation}
Using Eq.(4), there is
\begin{equation}
2\nu_{\triangle}^{2}(hh^{+})_{ij}=m_{1}^{2}\delta_{ij}+[V_{MNS}
diag(0,\triangle m^{2}_{21},\triangle m^{2}_{31})V_{MNS}^{+}]_{ij} ,
\end{equation}
where $\triangle m_{21}^{2}=m_{2}^{2}-m_{1}^{2}$ and $\triangle
m_{31}^{2}=m_{3}^{2}-m_{1}^{2}$. Since the sign of $\triangle
m_{31}^{2}$ is undetermined at present, distinct patterns for the
neutrino mass hierarchy are possible. $\triangle m_{31}^{2}>0$ is
referred to as normal hierarchy $(NH)$ with $m_{1}<m_{2}<m_{3}$ and
$\triangle m_{31}^{2}<0$ is defined as inverted hierarchy ($IH$)
with $m_{3}<m_{1}<m_{2}$. Combing Eq.(7) and Eq.(8), there is the
following relation

\begin{equation}
2\nu_{\triangle}^{2}\geq 1.96\times
10^{4}(\frac{200GeV}{m_{H}})^{2}[V_{MNS} diag (0,\triangle
m_{21}^{2}, \triangle m_{31}^{2}) V_{MNS}^{+}]_{e\mu}.
\end{equation}

In our following numerical estimation, we will consider this
constraint on the triplet $VEV$ $\nu_{\triangle}$ to calculate the
cross sections for single production of the doubly charged $Higgs$
boson $H^{--}$ via $e\gamma$ collision at the $ILC$ and $LHeC$.

\noindent{\bf 3. Single production of the doubly charged $\bf Higgs$
boson $H^{--}$ at the $\bf ILC$}

The $LHC$ can generate very massive new particles and will
essentially enlarge the possibilities of testing for new physics
effects, while the $ILC$ is also need to complement the probe of the
new particles with detailed measurement. For the $ILC$, the center
of mass ($c. m.$) energy $\sqrt{s}=0.5 - 1TeV$ and the typical
integrated luminosity $\pounds_{int}=0.5-1$ $ab^{-1}$ is currently
being designed [13, 14]. An unique feature of the $ILC$ is that it
can be transformed to $\gamma\gamma$ or $e\gamma$ collision with the
photon beam generated by laser-scatting method. The effective
luminosity and energy of the $\gamma\gamma$ and $e\gamma$ collisions
are expected to be comparable to those of the $ILC$. In some
scenarios, they are the best instrument for discovery of the new
physics signatures. In this section, we will consider single
production of the doubly charged $Higgs$ boson $H^{--}$ predicted by
the $HTM$ via $e\gamma$ collision.

\begin{figure}[htb]
\begin{center}
\epsfig{file=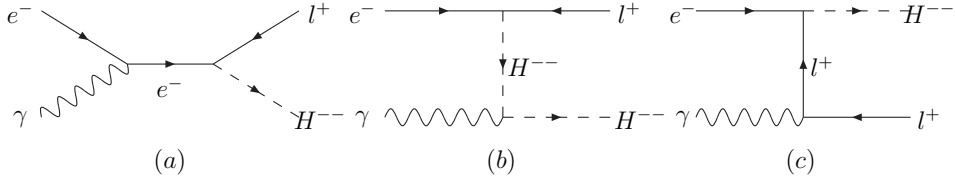,scale=0.8}
 \caption{The $Feynman$ diagrams  for the partonic process
 $e^{-}(p_{1})+\gamma(p_{2})\rightarrow \hspace*{1.8cm}
 l^{+}(p_{3})+H^{--}(p_{4})$ in the HTM.}
 \label{ee}
\end{center}
\end{figure}

From discussions given in section 2, we can see that the doubly
charged $Higgs$ boson $H^{--}$ can be produced via $e\gamma$
collision associated with a positive lepton. The relevant $Feynman$
diagrams are shown in Fig.1, in which $l^{+}$ denotes the lepton
$e^{+},\mu^{+}$ or $\tau^{+}$.
 For the process $e^{-}(p_{1})+\gamma(p_{2})\rightarrow
 l^{+}(p_{3})+H^{--}(p_{4})$, the renormalization amplitude can be
 written as
\begin{eqnarray}
M&=&M_{a}+M_{b}+M_{c}=-\frac{2eh_{el}}{(p_{1}+p_{2})^{2}}v^{T}(p_{3})C^{-1}P_{L}(\not{\hspace*{-0.15cm}p_{1}+
\not{\hspace*{-0.1cm}p_{2}}})\gamma^{\mu}u(p_{1})\varepsilon_{\mu}(p_{2})\nonumber\\
&&-\frac{4eh_{el}}{(p_{2}-p_{4})^{2}-m^{2}_{H}}v^{T}(p_{3})C^{-1}P_{L}
u(p_{1})(2p_{4}-p_{2})^{\mu}\varepsilon_{\mu}(p_{2})\nonumber\\
&&-\frac{2eh_{el}}{(p_{1}-p_{4})^{2}}v^{T}(p_{3})\gamma^{\mu}(\not{\hspace*{-0.15cm}
p_{1}-\not{\hspace*{-0.1cm}p_{4}}})C^{-1}P_{L}u(p_{1})\varepsilon_{\mu}(p_{2}).
\end{eqnarray}
Here $P_{L}=(1-\gamma_{5})/2$ is the right-handed projection
operator and $\varepsilon_{\mu}(p_{2})$ is the polarization vector
of the photon $\gamma$. The explicit expressions for the $Yukawa$
coupling constants $h_{ee}$, $h_{e\mu}$, and $h_{e\tau}$ can be
written as [8, 9]
\begin{eqnarray}
h_{ee}&=&\frac{1}{\sqrt{2}\nu_{\triangle}}[m_{1}e^{-i\varphi_{1}}c^{2}_{12}c^{2}_{13}
+m_{2}s^{2}_{12}c^{2}_{13}+m_{3}e^{i(2\delta-\varphi_{2})})s^{2}_{13},
\end{eqnarray}
\begin{eqnarray}
h_{e\mu}&=&\frac{1}{\sqrt{2}\nu_{\triangle}}[m_{1}e^{-i\varphi_{1}}c_{12}c_{13}
(-s_{12}c_{23}-e^{-i\delta}c_{12}s_{13}s_{23})+m_{2}s_{12}c_{13}(c_{12}c_{23}-\nonumber\\
&&e^{-i\delta}s_{12}s_{13}s_{23})+m_{3}e^{i(\delta-\varphi_{2})}s_{13}c_{13}s_{23}],
\end{eqnarray}
\begin{eqnarray}
h_{e\tau}&=&\frac{1}{\sqrt{2}\nu_{\triangle}}[m_{1}e^{-i\varphi_{1}}c_{12}c_{13}(s_{12}s_{23}-e^{-i\delta}c_{12}c_{23}s_{13})
+m_{2}c_{13}s_{12}(-c_{12}s_{23}-\nonumber\\
&&e^{-i\delta}s_{12}s_{13}c_{23})+m_{3}e^{i(\delta-\varphi_{2})}s_{13}c_{13}c_{23}].
\end{eqnarray}

\begin{figure}[htb] \vspace{-0.3cm}
\begin{center}
\includegraphics[width=225pt,height=185pt]{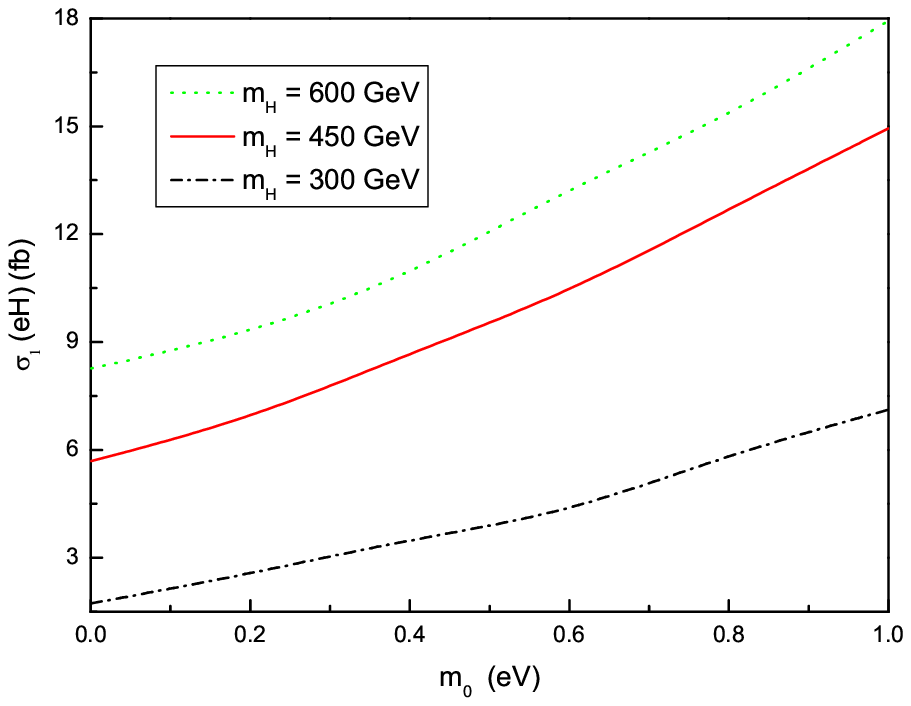} \put(-122,-1){
(a)}\put(90,-1){ (b)}
 \hspace{-0.8cm}\vspace{-0.25cm}
\includegraphics[width=225pt,height=185pt]{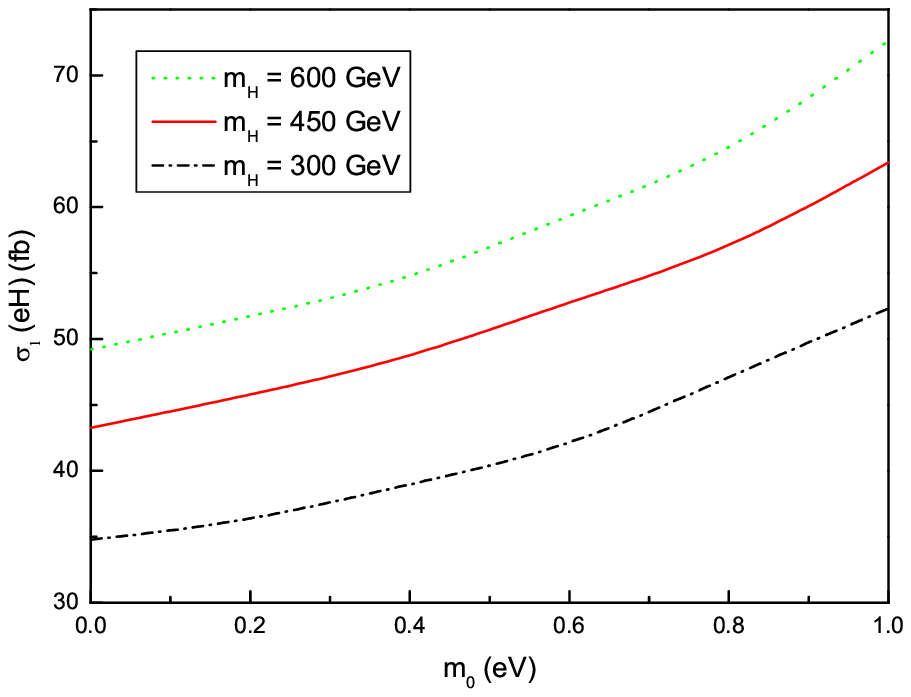} \hspace{-0.2cm}
 \caption{At the $ILC$, the production cross section $\sigma_{1}$ for the
 subprocess $e\gamma\rightarrow e^{+}H^{--}$ \hspace*{1.8cm}as a function $m_{0}$
  for different values of the $m_{H}$ in the $NH$$(a)$ and
  $IH$$(b)$ cases.}
 \label{ee}
\end{center}
\end{figure}

After calculating the cross section $\widehat{\sigma}(\widehat{s})$
for the subprocess $e\gamma\rightarrow l^{+}H^{--}$, the effective
cross section $\sigma_{1}(s)$ at the $ILC$ with the $ c. m.$ energy
$\sqrt{s}=1TeV$ can be obtained by folding the cross section
$\widehat{\sigma}(\widehat{s})$ with the photon distribution
function $f_{\gamma/e}(x)$ [21]
\begin{equation}
\sigma_{1}(s)=\int^{0.83}_{m_{H}^{2}/s}\widehat{\sigma}(\widehat{s})f_{\gamma/e}(x)dx,
\end{equation}
where $x=\widehat{s}/s$, in which $\sqrt{\widehat{s}}$ is the c.m.
energy of the subprocess $e^{-}\gamma\rightarrow l^{+}H^{--}$. In
above equation, we have neglected the mass of the lepton $l^{+}$ due
to $m_{H}>>m_{l^{+}}$.

\begin{figure}[htb] \vspace{-0.3cm}
\begin{center}
\includegraphics[width=225pt,height=185pt]{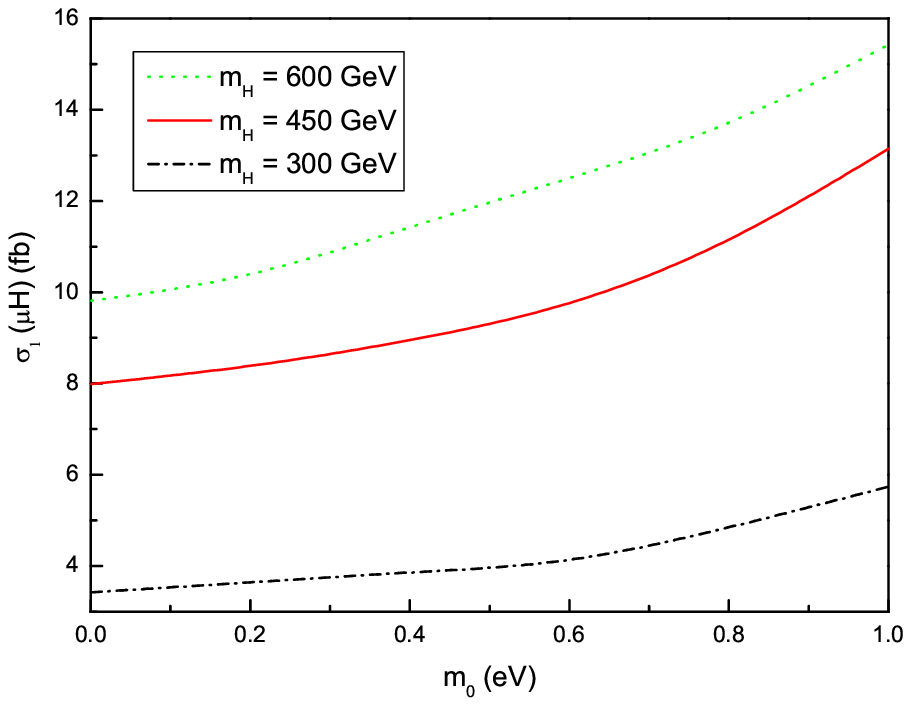} \put(-120,-1){
(a)}\put(90,-1){ (b)}
 \hspace{-0.8cm}\vspace{-0.25cm}
\includegraphics[width=225pt,height=185pt]{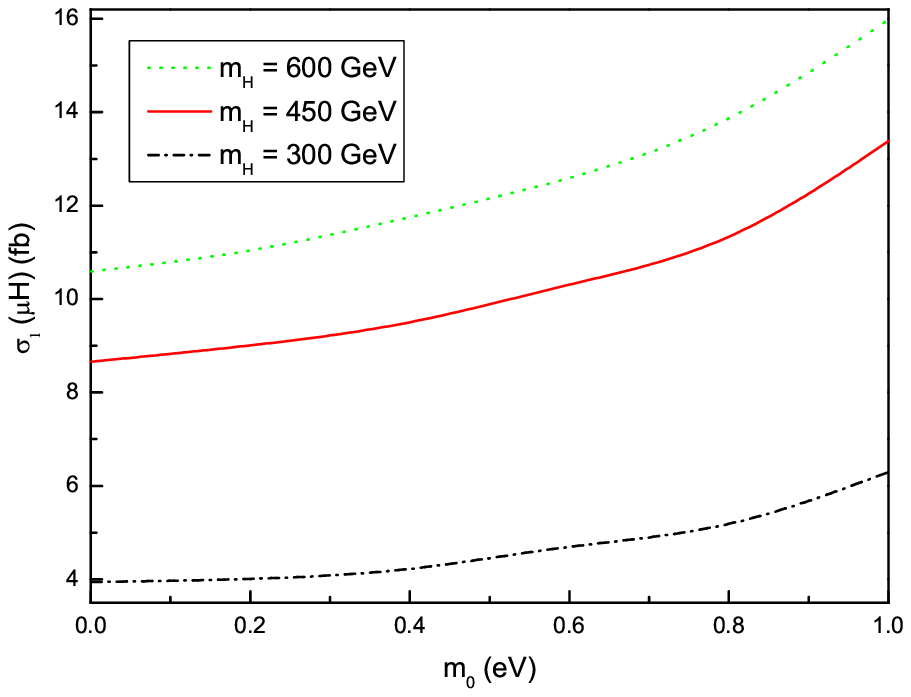} \hspace{-0.2cm}
 \caption{Same as Fig.2 but for the subprocess $e\gamma\rightarrow\mu^{+}H^{--}$.}
 \label{ee}
\end{center}
\end{figure}

It is obvious that the production cross section $\sigma_{1}$ is
dependent on the neutrino oscillation parameters, the triplet $VEV$
$\nu_{\triangle}$, and the mass parameter $ m_{H}$. Direct searches
for $H^{\pm\pm}$ have been performed at $LEP$ [22], Tevatron [23],
and $HERA$ [24] via the production mechanisms $e^{+}e^{-}\rightarrow
H^{++}H^{--}$ and $q\overline{q}\rightarrow H^{++}H^{--}$. Both of
these production mechanisms are depend on only one unknown parameter
$m_{H}$. The lower mass limits for $m_{H}$ from the Tevatron
searches are $m_{H}>110GeV\sim150GeV$ [23]. In this paper, we take
$m_{H}$ as a free parameter and assume that its value is in the
range of $300GeV\sim 600GeV$. The current experimental upper limit
for the $LFV$ process $\mu\rightarrow e\gamma$ has give severe
constraint on the triplet $VEV$ $\nu_{\triangle}$, as shown in
Eq.(9). In our numerical estimation, we take the minimal value given
by Eq.(9). In this case, the production cross section $\sigma_{1}$
is only dependent on $m_{H}$ and the neutrino oscillation
parameters.

 There are nine neutrino oscillation parameters: three neutrino
masses $m_{i}$ (or $m_{ij}$), three mixing angles $\theta_{ij}$, the
Dirac phase $\delta$ and two Majorana phases $\varphi_{1}$ and
$\varphi_{2}$. According the current constraints on the neutrinos
and mixing parameters from neutrino oscillation experiments [25], we
use the following values in this article
\begin{eqnarray}\nonumber
\hspace*{1.8cm}\triangle m_{21}^{2}&=& m_{2}^{2}-m_{1}^{2}\simeq
7.59\pm0.20 \left(\begin{array}{cc}+0.61\\-0.69
\end{array}\right)\times 10^{-5}eV^{2},\hspace*{0.8cm}
\end{eqnarray}
\begin{eqnarray}\nonumber
\hspace*{0.7cm}\triangle m_{31}^{2}&=& m_{3}^{2}-m_{1}^{2} =
-2.36\pm0.11(\pm0.37)\times 10^{-3}eV^{2} \hspace*{0.3cm}for \hspace*{0.3cm} IH,
\end{eqnarray}
\begin{eqnarray}
\hspace*{0.5cm}\triangle m_{31}^{2}&=& m_{3}^{2}-m_{1}^{2} =
2.46\pm0.12(\pm0.37)\times 10^{-3}eV^{2} \hspace*{0.3cm}for \hspace*{0.3cm} NH,
\end{eqnarray}
\begin{eqnarray}\nonumber
\hspace*{1cm}&&\theta_{12}=34.4\pm1.0\left(\begin{array}{cc}+3.2\\-2.9
\end{array}\right)^{o},\hspace*{0.5cm}
\theta_{23}=42.8^{+4.7}_{-2.9}\left(\begin{array}{cc}+10.7\\-7.3
\end{array}\right)^{o}, \\
&&\theta_{13}=5.6^{+3.0}_{-2.7}(\leq12.5)^{o}.
\end{eqnarray}
Information on the mass $m_{0}$ of the lightest neutrino, the Dirac
and Majorana phases can not be obtained from neutrino oscillation
experiments. To simply our paper, we will take $\delta=0$ and
$\varphi_{1}=\varphi_{2}=0$ to calculate the cross sections for
single production of $H^{--}$ in the cases of $NH$
 $(m_{0}=m_{1}<m_{2}<m_{3})$ and $IH$  $(m_{0}=m_{3}<m_{1}<m_{2})$. For
the free parameter $m_{0}$, we will assume its value in the range of
$0.001eV\leq m_{0}\leq 1eV$.

\begin{figure}[htb] \vspace{-0.3cm}
\begin{center}
\includegraphics[width=225pt,height=185pt]{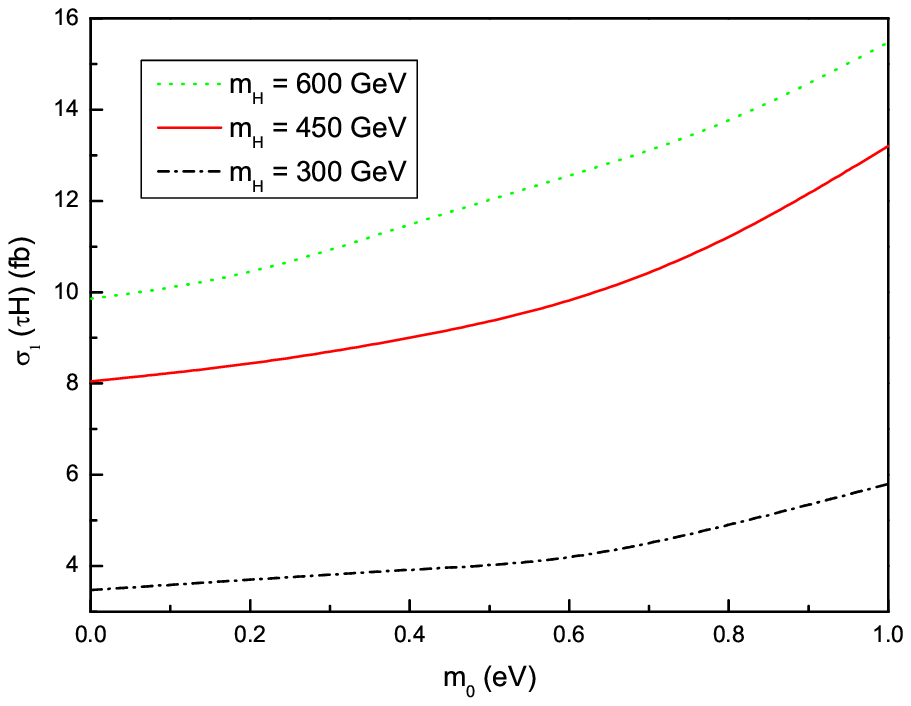} \put(-120,-1){
(a)}\put(90,-1){ (b)}
 \hspace{-0.8cm}\vspace{-0.25cm}
\includegraphics[width=225pt,height=185pt]{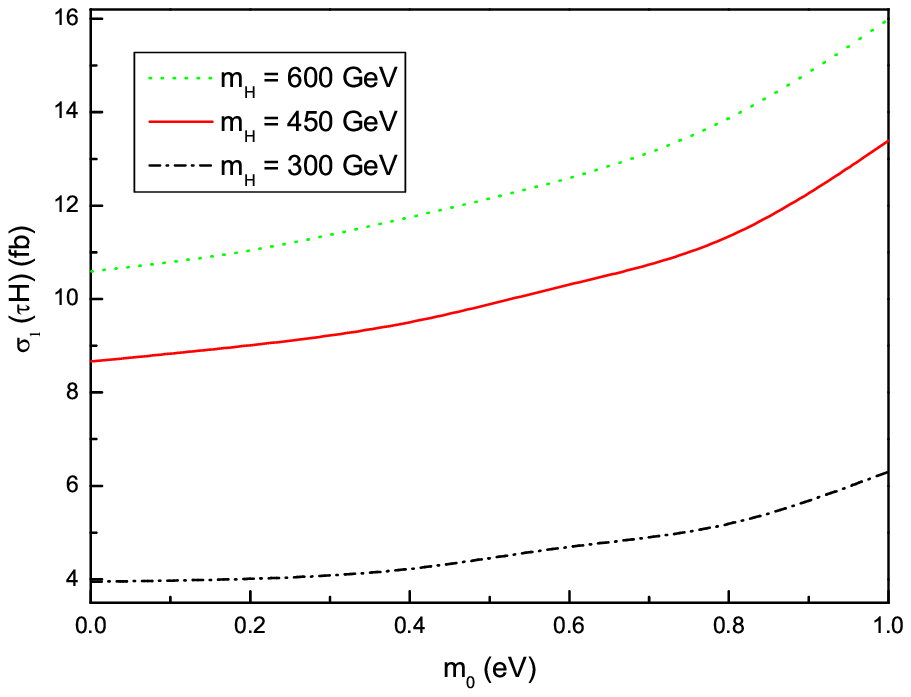} \hspace{-0.2cm}
 \caption{Same as Fig.2 but for the subprocess $e\gamma\rightarrow\tau^{+}H^{--}$. }
 \label{ee}
\end{center}
\end{figure}

From above discussions we can see that the production cross section
for the process $e\gamma\rightarrow l^{+}H^{--}$ depends on two free
parameters $m_{0}$ and $m_{H}$. In our numerical estimation, we take
the cross section $\sigma_{1}$ as a function of the lightest
neutrino mass $m_{0}$ for three values of $m_{H}$. Our numerical
results are summarized in Fig.2, Fig.3, and Fig.4, which correspond
the subprocesses $e\gamma\rightarrow e^{+}H^{--}$,
$e\gamma\rightarrow \mu^{+}H^{--}$, and $e\gamma\rightarrow
\tau^{+}H^{--}$, respectively. One can see from these figures that
the value of the production cross section $\sigma_{1}$ increases as
$m_{H}$ increases, which is because the minimal value of the triplet
$VEV$ $\nu_{\triangle}$ is proportional to the factor $1/m_{H}$. In
both of the NH and IH cases, the production cross section
$\sigma_{1}(\mu H)$ for the subprocess $e\gamma\rightarrow
\mu^{+}H^{--}$ is approximately equal to that for the subprocess
$e\gamma\rightarrow \tau^{+}H^{--}$ and their values are smaller
than $16fb$ in most of the parameter space of the $HTM$. This is due
to $h_{e\mu}\approx h_{e\tau}$, which reflects the neutrino mixing
patterns. However, for the subprocess $e\gamma\rightarrow
e^{+}H^{--}$, it is not this case. For the NH spectrum, its cross
section $\sigma_{1}(e H)$ is larger or smaller than the cross
section $\sigma_{1}(\mu H)$ ( or $\sigma_{1}(\tau H)$), which
depends on the mass $m_{0}$ of the lightest neutrino. For the IH
spectrum, the cross section $\sigma_{1}(e H)$ is always larger than
the cross section $\sigma_{1}(\mu H)$ ( or $\sigma_{1}(\tau H)$).
For $0.001eV\leq m_{0}\leq 1eV$ and $300GeV\leq m_{H}\leq 600GeV$,
the values of the cross sections for production of $H^{--}$
associated with a positron $e^{+}$ via $e\gamma$ collision at the
$ILC$ with $\sqrt{s}=1TeV$ are in the ranges of $1.7fb \sim 17.9fb$
and $34.8fb\sim 72.6fb$ corresponding to the $NH$ and $IH$ cases,
respectively.

\begin{figure}[htb] \vspace{-0.3cm}
\begin{center}
\includegraphics[width=225pt,height=185pt]{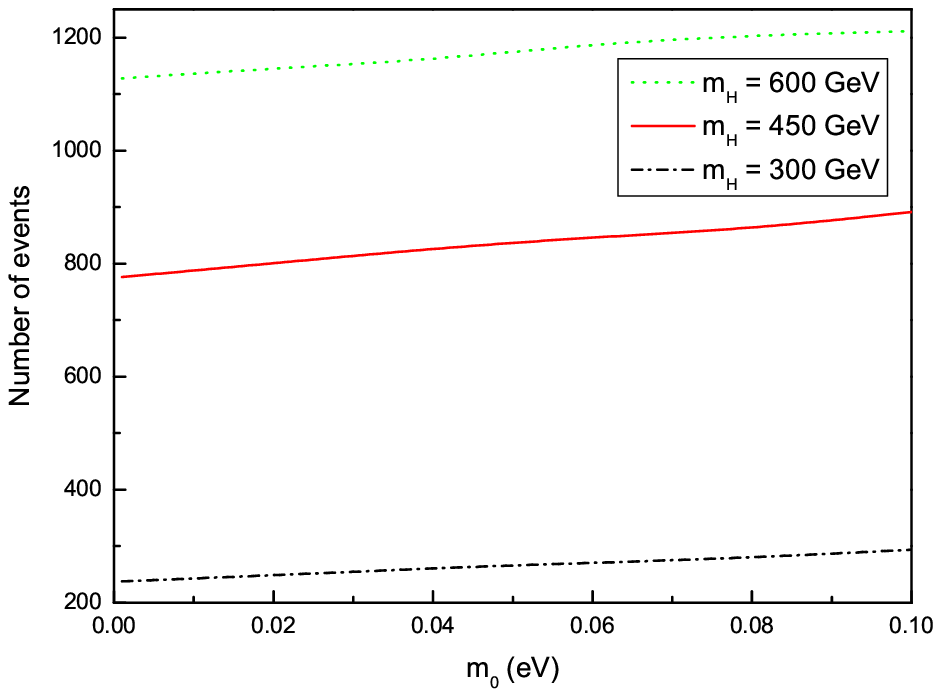} \put(-120,-1){
(a)}\put(90,-1){ (b)}
 \hspace{-0.8cm}\vspace{-0.25cm}
\includegraphics[width=225pt,height=185pt]{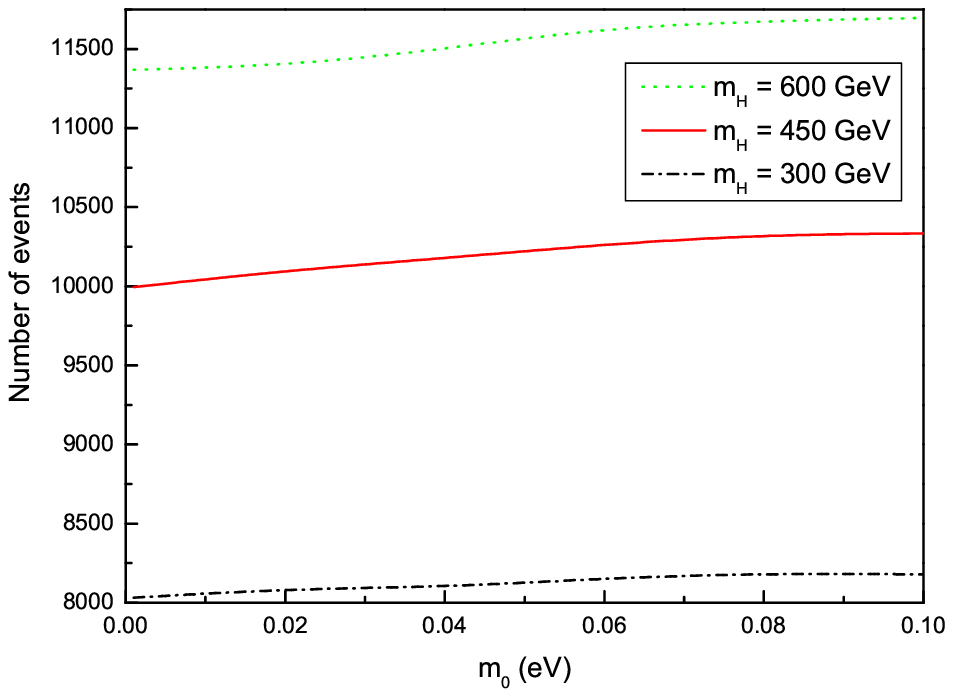} \hspace{-0.2cm}
 \caption{The number of the $\overline{e}\mu\mu$(a) and $\overline{e}ee$ (b)events at
the $ILC$.  }
 \label{ee}
\end{center}
\end{figure}

Ref. [9] has shown that, for $\nu_{\triangle}<0.1MeV$, the leptonic
decays of $H^{\pm\pm}$ are the dominant decay modes, while the decay
$H^{\pm\pm}\rightarrow W^{\pm}W^{\pm}$ is negligible. In the context
of the $HTM$, the leptonic decay channels $H^{\pm\pm}\rightarrow
l^{\pm}l^{\pm}$ have been extensively studied in Refs. [11, 26] and
the direct connections between their branching ratios and the
triplet $Yukawa$ couplings $h_{ij}$ (i.e. the neutrino oscillation
parameters) are discussed in these papers. For the leptonic decays
of $H^{\pm\pm}$, the possible decay modes are $ee$, $\mu\mu$,
$\tau\tau$, $e\mu$, $e\tau$ and $\mu\tau$. In the case of $\delta=0
$ and $\varphi_{1}=\varphi_{2}=0$, $\mu\mu$ and $\tau\tau$ are the
dominate decay modes, their branching ratios are roughly equal to
each other for the $NH$ spectrum and $m_{1}<0.1eV$. While for the
$IH$ mass spectrum and $0.001eV\leq m_{3}\leq 0.1eV$, the doubly
charged $Higgs$ boson $ H^{--}$  mainly decays to $ee$ and its
branching ratio is $32\%\leq Br(H^{--}\rightarrow ee)\leq 53\%$.
Thus, for the final states generated by the subprocess
$e\gamma\rightarrow e^{+}H^{--}$, we consider the three leptons
$\overline{e}\mu\mu$ and $\overline{e}ee$ for the $NH$ and $IH$
cases, respectively. The number of the $\overline{e}\mu\mu$ and
$\overline{e}ee$ events at the $ILC$ with $\sqrt{s}=1TeV$ and the
integrated luminosity $\pounds_{int}=500fb^{-1}$ are shown in Fig.5,
where we have taken $m_{0}$ in the range of $0.001eV\sim 0.1eV$.
From this figure, one can see that there will be several hundreds
and up to ten thousands three lepton events to be generated per year
in future $ILC$ experiments. It is obvious that this is not the
realistic number of the observed events. To take into account
detector acceptance we should impose appropriate cuts. For example,
the observed leptons must carry a minimum energy and respect a
suitable rapidity cut, the angles of the observed leptons relative
to the beam, $ \theta_{e}$, $ \theta_{\mu}$,  must be restricted to
some ranges [27]. Detailed analysis is needed, which is beyond the
scope of this paper.

It is well known that, at the $ILC$, the $LFV$ processes can provide
extremely clear signatures and are experimentally interesting. The
leptonic decays of the doubly charged $Higgs$ bosons give rise to
same-sign lepton pairs, which can generate distinct experimental
signals. The three lepton states $\overline{e}\mu\mu$ and
$\overline{e}ee$ generated by the subprocess $e\gamma\rightarrow
e^{+}H^{--}$ are almost free of the $SM$ backgrounds at the $ILC$,
which have been investigated in Ref. [27]. Thus, as long as the
triplet $VEV$ $\nu_{\triangle}$ is enough small, the doubly charged
$Higgs$ boson $H^{--}$ predicted by the $HTM$ might be detected in
future $ILC$ experiments.

\noindent{\bf 4. Single production of the doubly charged $\bf Higgs$
boson $H^{--}$ at the $\bf LHeC$}

Recently, the high-energy $ep$ collision has been considered at the
$LHC$, which is called the $LHeC$ [15]. For the $LHeC$, the energy
$E_{p}$ of the incoming proton is given by the $LHC$ beam, and the
energy $E_{e}$ of the incoming electron is in the range of
$50-200GeV$, corresponding to the $c. m.$ energies of
$\sqrt{s}=2\sqrt{E_{p}E_{e}}\approx 1.18-2.37TeV$. Its anticipated
integrated luminosity is at the order of $10-100fb^{-1}$ depending
on the energy of the incoming electron and the design. Several
studies have been performed to discuss the possibility of detecting
the $Higgs$ boson at the $LHeC$ [28]. In this section, we consider
single production of the doubly charged $Higgs$ boson $H^{--}$ via
$e\gamma$ collision at the $LHeC$.

It is well known that the equivalent photon approximation ($EPA$)
can be successfully to describe most of the processes involving
photon exchange [29]. A significant fraction of $pp$ collision at
the $LHC$ will involve quasi-real photon interactions occurring at
energies well beyond the electroweak energy scale. The quasi-real
photons emitted by the incoming protons have low virtuality $Q^{2}$
and scattered with small angles from the beam pipe. The photon
spectrum depending on virtuality $Q^{2}$ and energy $E_{\gamma}$ can
be described by the following relation [30, 31]
\begin{equation}
dN=\frac{\alpha}{\pi}\frac{dE_{\gamma}}{E_{\gamma}}\frac{dQ^{2}}{Q^{2}}
[(1-\frac{E_{\gamma}}{E_{p}})(1-\frac{Q^{2}_{min}}{Q^{2}})F_{E}+\frac{E_{\gamma}^{2}}{2E_{p}^{2}}F_{M}]
\end{equation}
with
\begin{equation}
Q^{2}_{min}=\frac{m_{p}^{2}E_{\gamma}^{2}}{E_{p}(E_{p}-E_{\gamma})},\hspace*{1.0cm}
F_{E}=\frac{4m_{p}^{2}G_{E}^{2}+Q^{2}G_{M}^{2}}{4m_{p}^{2}+Q^{2}},
\end{equation}
\begin{equation}
\hspace*{1.8cm}G_{E}^{2}=\frac{G_{M}^{2}}{\mu_{p}^{2}}=(1+\frac{Q^{2}}{Q^{2}_{0}})^{4},\hspace*{1.0cm}
F_{M}=G_{M}^{2},\hspace*{1.0cm} Q_{0}^{2}=0.71GeV^{2}.
\end{equation}
Where $E_{p}$ is the energy of the incoming proton which is related
to the photon energy by $E_{\gamma}=\xi E_{p}$ and $m_{p}$ is the
mass of the proton. Thus, $\xi$ is the proton momentum fraction
carried by the photon. The magnetic moment of the proton is
$\mu_{p}^{2}=7.78$, $F_E$ and $F_M$ are functions of the electric
and magnetic form factors. Electromagnetic form factors are steeply
falling as a function of $Q^{2}$. Then, the $Q^{2}$ integrated
photon flux can be written as

\begin{equation}
f(E_{\gamma})=\int
^{Q^{2}_{max}}_{Q^{2}_{min}}\frac{dN}{dE_{\gamma}dQ^{2}}dQ^{2},
\end{equation}
where $Q^{2}_{max}\approx 2-4GeV^{2}$. Since the contribution to the
above integral formula is very small for $Q^{2}_{max}>2GeV^{2}$, in
our numerical estimation, we will take $Q^{2}_{max}\approx2GeV^{2}$.

\begin{figure}[htb] \vspace{-0.3cm}
\begin{center}
\includegraphics[width=225pt,height=185pt]{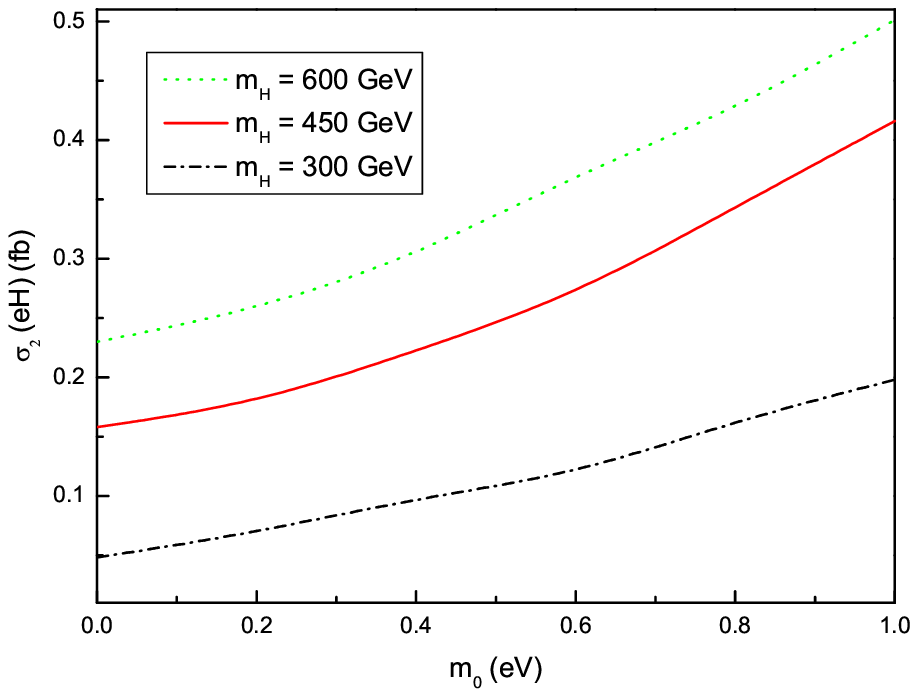} \put(-120,-1){
(a)}\put(90,-1){ (b)}
 \hspace{-0.8cm}\vspace{-0.25cm}
\includegraphics[width=225pt,height=185pt]{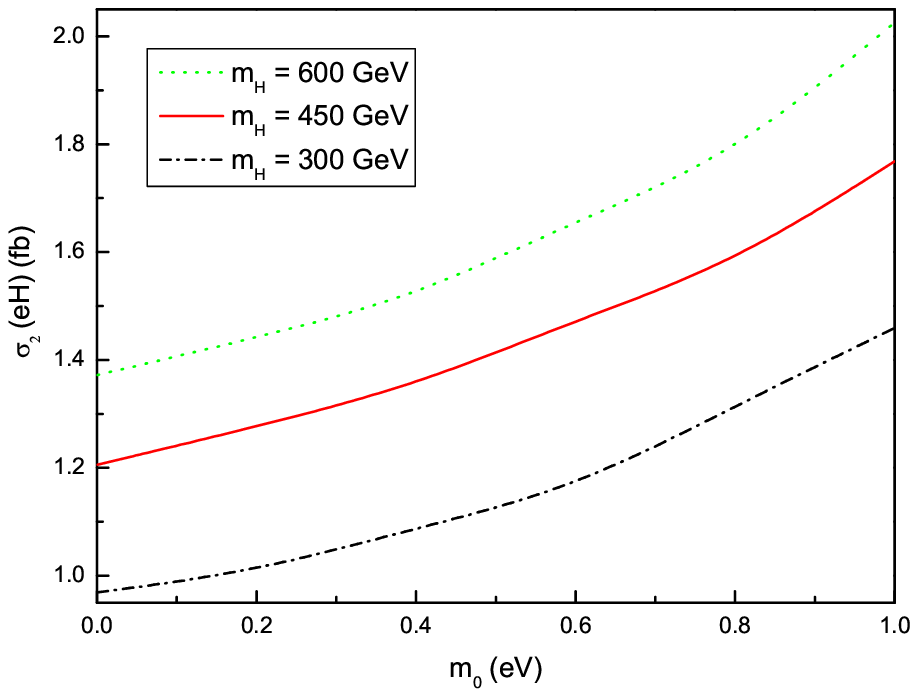} \hspace{-0.2cm}
 \caption{At the $LHeC$, the production cross section $\sigma_{2}$
for the subprocess $e\gamma\rightarrow e^{+}H^{--}$
\hspace*{1.8cm}as a function $m_{0}$ for different values of the
$m_{H}$ in the $NH$$(a)$ and $IH$$(b)$ cases. }
 \label{ee}
\end{center}
\end{figure}

\begin{figure}[htb] \vspace{-0.3cm}
\begin{center}
\includegraphics[width=225pt,height=185pt]{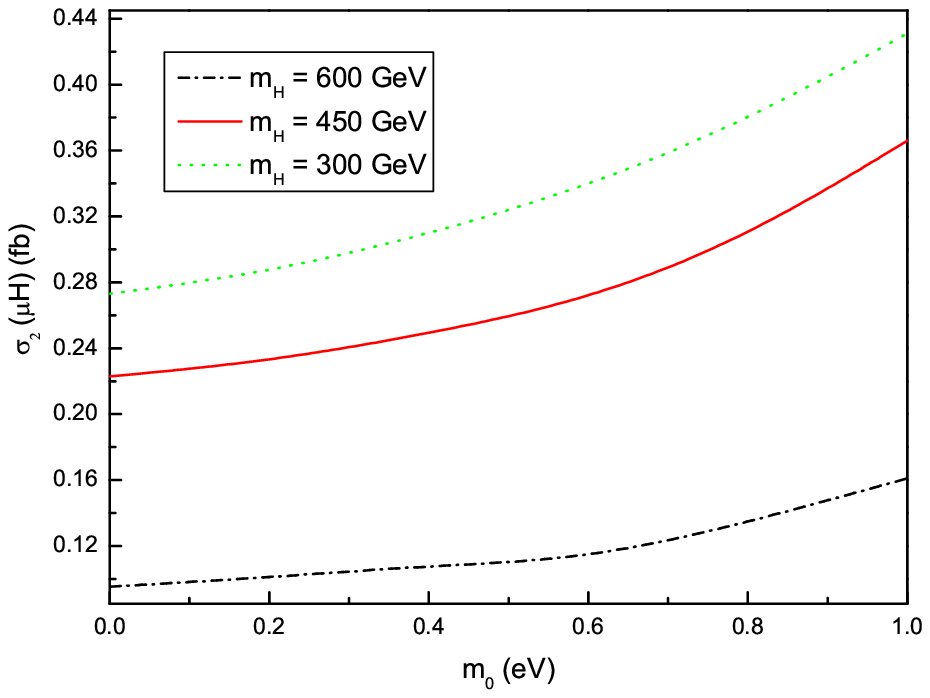} \put(-120,-1){
(a)}\put(90,-1){ (b)}
 \hspace{-0.8cm}\vspace{-0.25cm}
\includegraphics[width=225pt,height=185pt]{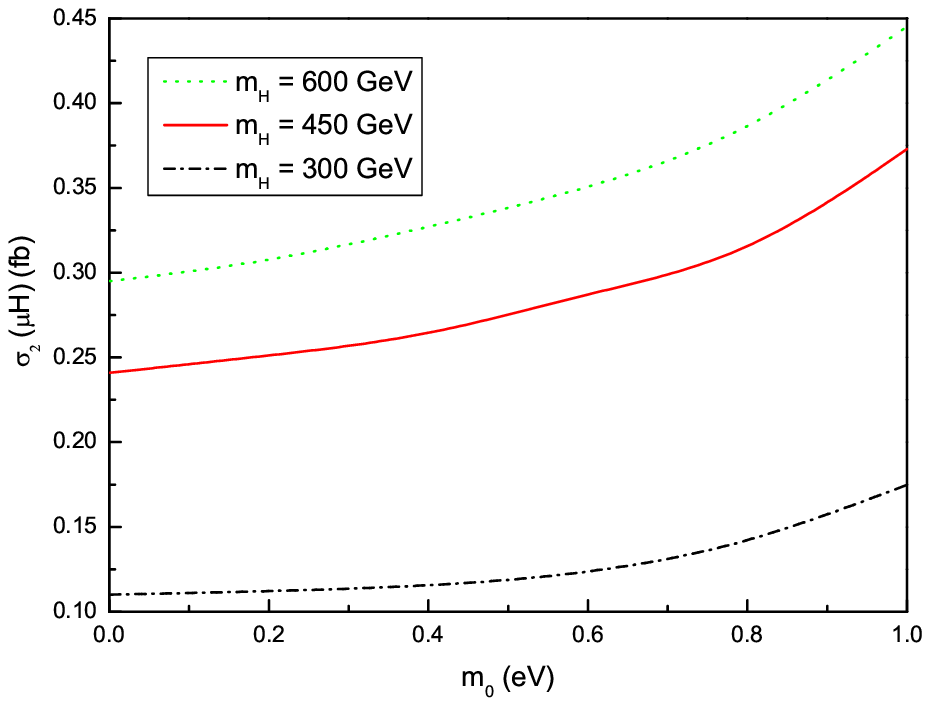} \hspace{-0.2cm}
 \caption{Same as Fig.6 but for the subprocess $e\gamma\rightarrow \mu^{+}H^{--}$. }
 \label{ee}
\end{center}
\end{figure}

At the $LHeC$, the effective production cross section
$\sigma_{2}(lH)$ for the subprocess $e\gamma\rightarrow l^{+}H^{--}$
can be written as

\begin{equation}
\sigma_{2}(lH)=\int^{\xi_{max}}_{(m_{H}+m_{l})^{2}/s}\sigma(\widehat{s})f(\xi
E_{p})d\xi,
\end{equation}
where $\widehat{s}=4E_{e}E_{\gamma}=\xi s$ with $E_{e}=140GeV$ and
$E_{p}=7TeV$. To detect the intact proton or its loss energy, the
forward proton detectors are needed. The $ATLAS$ forward physics
($AFP$) collaboration will have forward detectors with an acceptance
of $0.0015<\xi<0.15$ [31]. The acceptance of the $CMS-TOTEM$ forward
detectors can reach $0.0015<\xi<0.5$ [32]. In our numerical
estimation, we will take $\xi_{max}=0.5$ and
$\xi_{min}=(m_{H}+m_{l})^{2}/s$, which corresponds the $ c. m.$
energy of the subprocess $e\gamma\rightarrow l^{+}H^{--}$,
$\sqrt{\widehat{s}}\geq m_{H}+m_{l}$.

Our numerical results are summarized in Fig.6 and Fig.7, which plot
the production cross sections for the subprocesses
$e\gamma\rightarrow e^{+}H^{--}$ and $e\gamma\rightarrow
\mu^{+}H^{--}$ at the $LHeC$ as functions of the lightest neutrino
mass $m_{0}$ for three values of the mass $m_{H}$ of the doubly
charged $Higgs$ boson. In these figures we have shown our results
for the $NH$ and $IH$ cases. Our calculation shows that, at the
$LHeC$, the effective production cross section for the subprocess
$e\gamma\rightarrow\mu^{+}H^{--}$ is roughly equal to that for the
subprocess $e\gamma\rightarrow\tau^{+}H^{--}$, which is similar with
the conclusion at the $ILC$. Thus, we have not given the contours
for the subprocess $e\gamma\rightarrow\tau^{+}H^{--}$. From these
figures, one can see that the production cross section of the
subprocess $e\gamma\rightarrow\l^{+}H^{--}$ at the $LHeC$ is smaller
than that for the $ILC$. For $0.001eV\leq m_{0}\leq 1eV$ and
$300GeV\leq m_{H}\leq 600GeV$, the values of the production cross
section $\sigma_{2}(e H)$ and $\sigma_{2}(\mu H)$ are in the ranges
of $0.048fb\sim0.51fb$ and $0.095fb\sim0.43fb$, respectively, for
the $NH$ spectrum, while are in the ranges of $0.97fb\sim2.03fb$ and
$0.11fb\sim0.45fb$, respectively, for the $IH$ spectrum. If we
assume the integrated luminosity $L_{int}=100fb^{-1}$ of the $LHeC$
with $\sqrt{s}=1.96TeV$, which corresponds $E_{e}=140GeV$ and
$E_{p}=7TeV$, then there will be several and up to hundreds
$e^{+}H^{--}$events to be generated per year.

The process $ep\rightarrow e^{+}H^{--}+X$ with $H^{--}$ decaying to
$ee$ can give rise to number of signal events with same-sign
electron pair and an isolated positron, which is almost free of the
$SM$ backgrounds. Thus, it is also possible to detect the signatures
of the doubly charged $Higgs$ boson $H^{--}$ via its production
associated with a positron at the $LHeC$. Certainly, detailed
confirmation of the observability of the signals generated by the
$HTM$ at the $LHeC$ experiments would require Monte-Carlo
simulations of the signals and backgrounds, which is beyond the
scope of this paper.

\noindent{\bf5. Conclusions and discussions}

Doubly charged $Higgs$ bosons ($H^{\pm\pm}$) appear in some popular
new physics models beyond the $SM$, which can accommodate neutrino
masses. This kind of new particles have distinct experimental
signals through their decay to same-sign lepton pairs. Their
observation in future high energy collider experiments would be a
clear evidence of new physics. Thus, searching for $H^{\pm\pm}$ is
one of the main goals of current and future high energy collider
experiments.

The $HTM$ is one of the attractive new physics models, in which a
$Higgs$ triplet is introduced and the neutrino masses can be
obtained by the product of the triple $VEV$ $\nu_{\Delta}$ and the
triplet $Yukawa$ coupling $h_{ij}$. This model predicts the
existence of the doubly charged $Higgs$ bosons ($H^{\pm\pm}$). In
this paper, we consider production of $H^{--}$ associated with a
positive lepton $l^{+}$ via $e\gamma$ collision at the $ILC$ and
$LHeC$. The production cross sections are dependent on the neutrino
oscillation parameters, the triplet $VEV$ $\nu_{\triangle}$, and the
mass parameter $ m_{H}$. In our numerical estimations, we use the
lower bound on $\nu_{\Delta}^{2}$ given by the $LFV$ process
$\mu\rightarrow e\gamma$, assume $\delta=0$ and
$\varphi_{1}=\varphi_{2}=0$, and take the experimental measurement
values for the parameters $s_{ij}$, $c_{ij}$, and $m_{ij}$. In this
case, the production cross sections depend only on two free
parameters, the mass parameter $m_{H}$ and the lightest neutrino
mass $m_{0}$. From our numerical results, we can obtain following
conclusions.

(1) The cross section for production of the doubly charged $Higgs$
boson  $H^{--}$ associated with a positive lepton $l^{+}$ at the
$ILC$ is generally larger than that at the $LHeC$. The production
cross sections  for the $IH$ mass spectrum are larger than those for
the $NH$ mass spectrum at these two kinds of collider experiments.

(2) Because of $h_{e\mu}\simeq h_{e\tau}$, the production cross
section of the subprocess $e\gamma\rightarrow \mu^{+}H^{--}$ is
roughly equal to that of the subprocess
$e\gamma\rightarrow\tau^{+}H^{--}$ at both the $ILC$ and $LHeC$
experiments. For the $IH$ mass spectrum, the cross section
$\sigma(eH)$ is always larger than $\sigma(\mu H )$, while for the
$NH$ mass spectrum, $\sigma(eH)$ is larger or smaller than
$\sigma(\mu H )$ which depends on the value of the lightest neutrino
mass $m_{0}$.

(3) For $0.001eV\leq m_{0}\leq 1eV$ and $300GeV\leq m_{H}\leq
600GeV$, the values of the cross section $\sigma(eH)$ at the $ILC$
with $\sqrt{s}= 1 TeV$ are in the ranges of $1.7fb -- 17.9fb$ and
$34.8fb -- 72.6fb$ corresponding with the $NH$ and $IH$ cases,
respectively. While their values at the $LHeC$ with
$\sqrt{s}=1.97TeV$ are in the ranges of $0.048fb--0.51fb$ and
$0.97fb--2.03fb$ for the $NH$ and $IH$ cases, respectively.

(4) For the leptonic decays of $H^{--}$, the process
$e\gamma\rightarrow l^{+}H^{--}$ can produce three lepton final
state, which is almost free of the $SM$ backgrounds. In the context
of the $HTM$, this process can only generate number of three lepton
events at the $LHeC$, while, with reasonable values of the relevant
free parameters, it can give rise to large number of three lepton
events at the $ILC$. For instance, for $0.001eV\leq m_{0}\leq 0.1eV$
and $300GeV\leq m_{H}\leq 600GeV$, the subprocess
$e\gamma\rightarrow l^{+}H^{--}$ with $H^{--}$ decaying to $ee$ can
generate $8.0\times 10^{3} - 1.17\times 10^{4} $ $\overline{e}ee$
events at the
 $ILC$ experiment with $\sqrt{s}=1000GeV$ and the integrated luminosity
$\pounds_{int}=500fb^{-1}$ in the $IH$ mass spectrum. Thus, the
possible signals of  $H^{--}$ predicted the $HTM$ might be detected
via this process in the future $ILC$ experiments.

In our numerical estimation, we have taken the minimal value of the
triplet $VEV$ $\nu_{\Delta}$ demanded by the current upper
experimental bound for the $LFV$ process $\mu\rightarrow e\gamma$.
However, the constraint on the $HTM$ from the process
$\mu\rightarrow ee\overline{e}$ is generally stronger than that from
the process $\mu\rightarrow e\gamma$. For example, Ref. [12] has
shown that, for $\varphi_{1}=\varphi_{2}=0$ and $
\sin^{2}2\theta_{13}=0$, the processes $\mu\rightarrow
ee\overline{e}$ and $\mu\rightarrow e\gamma$ demand
$\nu_{\Delta}m_{H}\geq400eV\cdot GeV $ and
$\nu_{\Delta}m_{H}\geq100eV\cdot GeV $, respectively. It is obvious
that the minimal value of $\nu_{\Delta}$ demanded by $\mu\rightarrow
e\gamma$ is smaller than that for $\mu\rightarrow ee\overline{e}$.
Thus, if we use the constraints on the free parameters of the  $HTM$
coming from the process $\mu\rightarrow ee\overline{e}$, the
production cross sections and the number of the signal events will
be decreased at least by one order of magnitude. However, as long as
$\nu_{\Delta}<10^{-5}GeV$, there will be several $ \overline{e}ee$
events to be generated at the $ILC$.

It has been shown [26] that the  Majorana phases $\varphi_{1}$ and
$\varphi_{2}$ have large effects on the branching ratios
$Br(H^{\pm\pm}\rightarrow l_{i}^{\pm}l_{j}^{\pm})$. Thus,
$\varphi_{1}$ and $\varphi_{2}$ also have effects on the cross
section for production of the doubly charged $Higgs$ boson  $H^{--}$
associated with a positive lepton $l^{+}$. If we vary the values of
$\varphi_{1}$ and $\varphi_{2}$, the numerical results for the cross
sections $\sigma(eH)$, $\sigma(\mu H )$, and $\sigma(\tau H)$ are
also changed. However, our physical conclusions are not changed.

Certainly, the doubly charged $Higgs$ boson $H^{++}$ can also  be
singly produced at the $ILC$ and the $LHeC$ via the
charge-conjugation processes of the corresponding processes for
$H^{--}$. Similar to above calculation, we can give the values of
the production cross sections for $H^{++}$. Thus, the conclusions
for $H^{--}$ also apply to $H^{++}$.

If the doubly charged $Higgs$ bosons $H^{\pm\pm}$ are indeed found
at the $LHC$, various consistency checks will have to be performed.
One of the important tasks  is to exact calculate and  precise
measure the branching ratios $Br(H^{\pm\pm}\rightarrow
l_{i}^{\pm}l_{j}^{\pm})$ and see whether their values are consistent
with a neutrino mass matrix from oscillation data. We expect that
our work will be helpful to test the Higgs triplet mechanism for
neutrino masses and further to distinguish different new physics
models.

\vspace{0.5cm}
\section*{Acknowledgments} \hspace{5mm}
 This work is supported in part by the National Natural Science
Foundation of China under Grant No.10975067, the Specialized
Research Fund for the Doctoral Program of Higher Education(SRFDP)
(No.200801650002).

\end{document}